# Giant Magnetocaloric Effect in a High-Spin Shastry-Sutherland Dipolar Magnet

Jianjian Gong[1,2,*], Junsen Wang[3,4,*], Junsen Xiang[5,*], Zhaojun Mo[1,2,†], Lei Zhang[1], Xinyang Liu[4,5], Xuetong He[1,2], Lu Tian[1], Zhixing Ye[1], Huicai Xie[1], Xucai Kan[6], Xinqiang Gao[1,2], Zhenxing Li[7], Peijie Sun[5], Shouguo Wang[6], Wei Li[4,‡], Baogen Shen[1,2,5,6], and Jun Shen[1,2,7,§]

[1] Key Laboratory of Rare Earths, Ganjiang Innovation Academy, Chinese Academy of Sciences, Ganzhou 341119, China
[2] School of Rare Earths, University of Science and Technology of China, Hefei 230026, China
[3] Anhui Province Key Laboratory of Condensed Matter Physics at Extreme Conditions, High Magnetic Field Laboratory, Chinese Academy of Sciences, Hefei 230031, China
[4] Institute of Theoretical Physics, Chinese Academy of Sciences, Beijing 100190, China
[5] Beijing National Laboratory for Condensed Matter Physics, Institute of Physics, Chinese Academy of Sciences, Beijing 100190, China
[6] Anhui Provincial Key Laboratory of Magnetic Functional Materials and Devices, School of Materials Science and Engineering, Anhui University, Hefei 230601, China
[7] Department of Energy and Power Engineering, School of Mechanical Engineering, Beijing Institute of Technology, Beijing 100081, China

Jianjian Gong, Junsen Wang, and Junsen Xiang contributed equally to this work.

Corresponding Authors: Zhaojun Mo, Wei Li, Jun Shen.

**Abstract:** The Shastry-Sutherland lattice is a prototypical frustrated quantum magnet. It is notable for its exactly solvable dimer-singlet ground state and hosts a wealth of magnetic phenomena under external fields. Here, this work investigates the high-spin ($S = 7/2$) Eu-based magnet $Eu_2MgSi_2O_7$ (EMSO) using low-temperature magnetothermal measurements and Monte Carlo simulations, revealing a giant magnetocaloric effect (MCE) in this Shastry-Sutherland compound. The entropy change peak value is found to be 55.0 J·kg$^{-1}$·K$^{-1}$ under a field change of $\mu_0 H = 0$–4 T, approximately 1.5 times larger than the commercial $Gd_3Ga_5O_{12}$ (GGG). Adiabatic demagnetization refrigeration achieves a lowest temperature of 151 mK, deeply into the sub-Kelvin regime. Furthermore, a distinctive cooling effect persists below about 1 T, a characteristic absent for conventional magnetic coolants. A dipolar Shastry-Sutherland model is introduced as a minimal model to describe this system; in particular, the experimentally revealed 1/3 magnetization pseudo-plateau can be ascribed to the presence of dipolar couplings between $Eu^{2+}$ ions, further stabilized by the thermal fluctuations, explaining the persistent cooling effect. This work establishes EMSO as a novel platform for exploring the dipolar Shastry-Sutherland system and for sub-Kelvin adiabatic demagnetization refrigeration.

## 1. Introduction

Geometrically frustrated magnetism, characterized by competing interactions that prevent conventional ordering, serves as a fertile ground for exotic spin states, unconventional phase transitions, and various emergent phenomena.[1,2] The renowned Shastry-Sutherland (SS) system,[3] with lattice geometry shown in **Figure 1**b, is such a paradigmatic example that has been investigated for decades. For the spin-1/2 case with proper couplings, it can have an exact soluble singlet-dimer ground state.[3] Intriguingly, it is found recently that there may exist a possible quantum spin liquid phase[4,5] and/or a deconfined quantum critical point[6] in this model. Furthermore, rich sequence of magnetization plateaus — explained as the crystal of triplons or triplon bound states[7] — as well as different spin supersolid states are observed under high fields.[8,9] The most studied experimental realization of the SS model is the layered compound $SrCu_2(BO_3)_2$,[10] which exhibits various spin states and phase transitions under magnetic field and pressure.[11,12] More recently, new rare-earth magnets with such highly frustrated lattice have emerged, opening a fresh avenue for exploring Shastry-Sutherland physics. These include the families $BaRE_2TCh_5$ (T = transition metal, Ch = O, S) with RE = La-Eu,[13-16] and $RE_2Be_2YO_7$ (Y = Ge, Si) with RE = Nd, Sm, Gd-Yb.[17-20] The properties of these systems await comprehensive experimental and theoretical investigations.

In recent years, frustrated magnets have also emerged as a promising alternative to hydrated paramagnetic salts for adiabatic demagnetization refrigeration (ADR),[21] since they possess enhanced low-temperature entropy due to spin frustration[22-26] and large cooling power because of higher magnetic density.[23,27] It is thus interesting to explore various geometrically frustrated lattice systems like the SS lattice here. So far, there are only limited studies of magnetocaloric and barocaloric effect in spin-1/2 SS magnets;[9,19] however, higher spin SS magnets have rarely been examined before. In viewing that large local spin increases the entropy storage capacity and also improves the hold time, it is expected that high spin SS system also give rise to better cooling performance.

In this work, we study an $S$ = 7/2 rare-earth magnet $Eu_2MgSi_2O_7$ (EMSO) with a SS lattice. As the $Eu^{2+}$ ions carry large magnetic moments, the dipolar interactions, which make the system more frustrated inherently, also play an essential role in determining the magnetic properties.[28] This system is thus dubbed as a dipolar SS (DSS) magnet. We conducted low-temperature magnetic, thermodynamic and magnetocaloric measurements on polycrystalline EMSO sample, and reveal its great potential as an ideal sub-Kelvin refrigerant. Notably, we report the first observation of a 1/3 magnetization pseudo-plateau in a DSS magnet. From Monte Carlo (MC) simulations of the corresponding model system, we reproduce this pseudo-plateau and reveal its mechanism as the interplay of entropy selection and dipolar interactions. In the magnetocaloric measurements, EMSO exhibits maximum magnetic entropy change $-\Delta S_M^{max}$ = 55.0 J·kg$^{-1}$·K$^{-1}$ under a moderate field variation of 4 T. From an initial condition of 1.8 K and 8 T, the lowest attained temperature is 151 mK. These results significantly outperform the commercial $Gd_3Ga_5O_{12}$ (GGG) in both achievable temperature and entropy change. Our work therefore establishes a new platform for investigating exotic magnetic states in high-spin DSS systems while also advancing practical applications for sub-Kelvin refrigeration.

## 2. Results

### 2.1. Rare-Earth Compound $Eu_2MgSi_2O_7$ with SS Lattice

We prepare polycrystalline samples of EMSO via solid-state reaction. As shown in **Figure 1**a, magnetic $Eu^{2+}$ ions with half-full $4f^7$ electron configuration ($L$ = 0, $S$ = 7/2) form the SS lattice in the ab-plane (**Figure 1**b). Its magnetic properties have not been investigated so far.

Within the 2D layer, the Eu-Eu nearest neighbor (NN) distance $d_1$ = 3.727 Å and the next NN distance $d_2$ = 4.227 Å, which are smaller than the interlayer Eu-Eu distance $c$ = 5.166 Å, in accordance with a previous report.[29] Such a crystal structure suggests that EMSO can be considered as a quasi-two-dimensional SS lattice system, with relatively high magnetic ion density of 12.1 nm$^{-3}$.

As a DSS lattice system, the compound EMSO features two dominant interactions: competing Heisenberg exchanges (intra-dimer $J'$ > 0 and inter-dimer $J$ > 0), and magnetic dipolar coupling. The frustrated $J$-$J'$ couplings are indicated in Figure 1b. The dipolar energy scale is $J_{dp} = \mu_0 \mu_{eff}^2 / 4\pi d_1^3 \approx$ 588 mK, with $\mu_{eff} \approx$ 7 $\mu_B$, according to the distance between adjacent Eu$^{2+}$ ions, which is comparable to the Heisenberg exchange (see spin model analysis below).

## 2.2. Magnetic and Thermodynamics Measurements

We conduct low-temperature magnetic and thermodynamics measurements on the EMSO compound. Figure 1c shows the specific heat ($C_p/T$) measured under different magnetic fields at temperatures as low as 150 mK. The zero-field specific heat displays a sharp peak at around 700 mK, signaling a magnetic phase transition. As the magnetic field increases, the peak progressively shifts to lower temperatures with modest amplitude reduction. At higher fields, the sharp peak transforms into a broad hump that moves toward elevated temperatures. Notably, the zero-field $C_p/T$ curve reveals shoulder-like features accompanying the main peak, forming a mountain-like profile. These features surpass those of commercial GGG refrigerant[30] with larger low-$T$ specific heat and thus stronger spin fluctuations [see Figure 1c], indicating superior potential of EMSO for magnetocaloric applications.

In Figure 1d, we show the results of magnetic entropy change, $-\Delta S_M \equiv S(H = 0)-S(H)$, calculated from the measured isothermal magnetization data. Under field change of $\mu_0 H$ = 0–2 T and 0–4 T, the $-\Delta S_M^{max}$ values are 36.7 J·kg$^{-1}$·K$^{-1}$ and 55.0 J·kg$^{-1}$·K$^{-1}$, respectively. In particular, the $-\Delta S_M^{max}$ value of EMSO at $\mu_0 H$ = 0–2 T is comparable to that of the commercial magnetic refrigerant GGG (36.0 J·kg$^{-1}$·K$^{-1}$),[31] but requires only half the field change (2 T vs. 4 T for GGG). For $\mu_0 H$ = 0–4 T, EMSO shows a significantly larger $-\Delta S_M$, 1.5 times the value of GGG under the same field change. Superior magnetocaloric performance at lower fields can effectively reduce application costs and enable refrigerator compactness. Furthermore, the $-\Delta S_M(T)$ curves under $\mu_0 H$ = 0–4 T calculated from $C_p(T)$ and $M(\mu_0 H)$ data show a very good agreement (see Supporting Information Figure S4).

## 2.3. Magnetocaloric Measurements and Persistent Cooling

We also conduct quasi-adiabatic demagnetization cooling measurements, as illustrated in Figure 1e. It is found that EMSO can reach a remarkably low temperature of 151 mK under the initial conditions of $T_i$ = 1.8 K and $\mu_0 H_i$ = 8 T. Notably, the lowest temperature in the isentropic curves of EMSO is observed not at 0 T but at $\mu_0 H_c \approx$ 1.13 T. This suggests the presence of a quantum critical point (QCP) at $\mu_0 H_c$. In addition, the isentropic curve exhibits a plateau below $\mu_0 H_c$, where the cooling temperature remains at a low value.

Compared EMSO with the commercial refrigerants GGG and LiGdF$_4$. Starting from the same initial condition ($T_i$ = 2 K and $\mu_0 H_i$ = 6 T, see Supporting Information Figure S5), we find

EMSO reaches 236 mK, lower than GGG (322 mK) and LiGdF$_4$ (480 mK).[32] The magnetic Grüneisen ratio $\Gamma_B$ is often used to characterize the universal cooling behaviors near the QCP.[33-37] Figure 1f reveals a peak-dip structure with a sign reversal, providing additional evidence for a possible field-driven QCP. Remarkably, even in polycrystalline form, EMSO demonstrates Grüneisen ratios exceeding those of GGG — despite both systems featuring high-spin ($S = 7/2$) ions. Overall, EMSO demonstrates exceptional magnetic refrigeration performance, combining ultralow base temperature, singular Grüneisen ratios, sustained cooling capability, and large entropy change (Figure 1d). These characteristics establish EMSO as a superior candidate for sub-Kelvin cooling applications.

### 2.4. 1/3 Magnetization Pseudo-Plateau

Figure 2a displays the field-cooled magnetic susceptibility ($\chi$) measured from 0.6 K to 2.0 K under an applied field of 0.01T. Above 1 K, the magnetic susceptibility curve shows a paramagnetic (PM) behavior. The Curie-Weiss fitting is performed in the low temperature PM region, which results in a Curie-Weiss temperature $\theta_{CW} \approx -1.65$ K and an effective magnetic moment $\mu_{eff} \approx 7.66$ $\mu_B$. The negative $\theta_{CW}$ value indicates the dominant antiferromagnetic (AFM) interactions between the Eu$^{2+}$ ions. Figure 2b shows the measured isothermal magnetization $M(\mu_0H)$ curves, under temperatures ranging from 0.4 K to 1.8 K. The inset shows the normalized magnetization ratio $R_M \equiv M(T)/M(T = 0.4 \text{ K})$. It is found that there is a crossing point at around $(R_M, \mu_0H_p) = (1.0, 0.5 \text{ T})$. For curves with temperature $T = 0.5$–0.8 K and fields smaller (larger) than 1 T, the corresponding $R_M$ is smaller (larger) than 1. This behavior reveals the emergence of a 1/3 magnetization pseudo-plateau around the crossing point.

Recall that in Figure 1c, EMSO exhibits pronounced $C_p/T$ values below 1.0 K and for fields under 1 T, reflecting its substantial magnetic entropy ($S_M$) in this regime. This demonstrates that the 1/3-magnetization pseudo-plateau directly correlates with enhanced magnetic entropy. In fact, for the classical SS model, it has been theoretically proposed that a magnetization pseudo-plateau appearing at $M/M_{sat} = 1/3$ is selected by thermal fluctuations.[38] Here, it is called "pseudo" in the sense that this plateau is not completely flat, and its defining feature is visualized by examining the differential magnetic susceptibility $\chi = dM/d(\mu_0H)$, which shows a peak-dip-peak behavior near this pseudo-plateau.[38]

In Figure 2c and d, we show the experimental data of isothermal magnetization and also the differential magnetic susceptibility of EMSO. Surprisingly, at low temperature ($T = 0.4$ K), there is indeed a peak-dip-peak structure at around $\mu_0H_p \approx 0.5$ T (as indicated by the red vertical dashed line), corresponding to the presence of a 1/3 magnetization pseudo-plateau. However, at elevated temperatures (e.g., 1.8 K, Figure 2d), such feature disappears, with no discernible signature remaining in either the magnetization or susceptibility measurements. These observations further indicate that thermal fluctuations play a dual role in the pseudo-plateau formation — they establish the pseudo-plateau at low temperatures yet destroy it upon heating. To our knowledge, this represents the first experimental observation of a 1/3 pseudo-plateau state in high-spin SS lattice, stabilized by an order-by-thermal-disorder mechanism.

### 2.5. Dipolar Shastry-Sutherland Model

To comprehend the experimental findings, we develop a minimal model for EMSO incorporating both spin exchange and long-range dipolar interactions. Monte Carlo simulations are performed, with the full dipolar terms treated using the Ewald summation method.[28,39-41] The magnetic behavior of EMSO is described by the following dipolar-Heisenberg Hamiltonian:

$$H = J\sum_{\langle i,j \rangle} \mathbf{S}_i \cdot \mathbf{S}_j + J'\sum_{\langle\langle i,j \rangle\rangle} \mathbf{S}_i \cdot \mathbf{S}_j + \frac{1}{2}\sum_{i,j}\sum_{u,v} D_{ij}^{uv} S_i^u S_j^v \quad (1)$$

which is dubbed the DSS model hereafter. Here $\langle i,j \rangle$ and $\langle\langle i,j \rangle\rangle$ represent inter- and intra-dimer interactions on the 2D SS lattice. The spin $S$ is treated as a classical vector of unit length. The dipolar interaction tensor takes the form $D_{ij}^{\mu\nu} = J_{dp}\frac{\delta^{\mu\nu} - 3e_{ij}^{\mu}e_{ij}^{\nu}}{r_{ij}^3}$, with $J_{dp} = \mu_0\mu_{eff}^2/4\pi d_1^3 \approx$ 588 mK, and $r_{ij}$ the distance connecting site $i$ and $j$, expressed in the unit of the intra-dimer lattice spacing $d_1$. We also consider the system coupled to magnetic field along $\hat{z}$ direction, by adding a term $-\mu_{eff}^2\mu_0 H \sum_i S_i^z$ to the Hamiltonian **Equation 1**.

MC simulations are conducted on a system comprising 12×12 unit cells, with each unit cell containing four lattice sites. By comparing numerically obtained magnetization curves with the experimental ones, we determine the coupling strength $J \approx 499$ mK and $J' \approx 997$ mK. Thus the dipolar interaction in this system turns out to be comparable with the Heisenberg interactions, and it also plays an essential role. The experimental and theoretical results are shown and compared in Figure 2c-f. In particular, our MC simulations show that the ground state of EMSO is a magnetically ordered state, called the orthogonal state, as illustrated in **Figure 3**b. Note this orthogonal state has previously been found in the square lattice dipolar magnet, via an order by thermal disorder mechanism.[42-45] As a comparison, phase diagram of the classical SS model without dipolar interaction is shown in Figure 3a, where the ground state is found to be the Y state in the small field regime.[38] Hence the magnetic state alters significantly by including a considerable dipolar interaction.

We further simulate the magnetization process for cases with and without dipolar interactions at even lower temperatures than experiments. Remarkably, it is found that the inclusion of dipolar interaction significantly enhances the flatness of this 1/3 magnetization pseudo-plateau (see Figure 3c). In fact, the pseudo-plateau of the original SS model becomes a bona fide plateau, in the sense that there is indeed a virtually flat plateau in the magnetization curve of the DSS model at low enough temperatures. Note for both cases, the nature of this plateau state is always an up-up-down (UUD) state, as illustrated in Figure 3a and b. In particular, the regime of the UUD state becomes larger by including the dipolar interaction. Moreover, by numerically minimizing the ground state energy of the DSS model, we find that there is a finite field regime where the ground state is the UUD state, in contrast to the original classical SS model, where the UUD phase vanishes at zero temperature.[46] By further increasing field, the DSS system enters the canted orthogonal state before it gets fully polarized. Here the system, besides forming an orthogonal structure in the *x-y* plane, also has a non-zero component along the *z* direction; therefore forms a canted structure.

## 3. Discussion

Combining experimental and theoretical studies, we present the first experimental evidence of an entropy-driven magnetization pseudo-plateau in high-spin DSS magnets, confirming the long-predicted order-by-thermal-disorder phenomenon. This strong entropic effect also impacts magnetic cooling behavior. The concurrent optimization of high magnetic entropy density and ultra-low ordering temperature remains a fundamental challenge in designing advanced sub-Kelvin refrigerants. As shown in **Figure 4**, the magnetocaloric performance of EMSO significantly outperformed that of both commercial GGG and most previously reported materials.[19,25,31,37,47-51] Furthermore, compared to the spin-1/2 SS lattice Yb$_2$Be$_2$GeO$_7$,[19] the high-spin ($S$ = 7/2) SS lattice ESMO exhibits a larger magnetic entropy change, with a value of 60.1 J·kg$^{-1}$·K$^{-1}$ (297.9 mJ·cm$^{-3}$·K$^{-1}$), and attains a cooling temperature of 151 mK. Our results demonstrate that the high-spin DSS magnet EMSO emerges as a promising frustrated material candidate to address the challenge mentioned above.

Specifically, the measured isentropic lines are rather flat below 1 T (Figure 1e), which is also seen in the negligibly small magnetic Grüneisen ratio $\Gamma_B$ for small fields (Figure 1f). We emphasize that two prominent valley-like regimes was observed in the triangular-lattice compound Na$_2$BaCo(PO$_4$)$_2$, where low temperatures are maintained by the strongly fluctuating spin supersolid phase.[26,52,53] In the DSS magnet EMSO, the observed persistent cooling effect arises from substantial spin fluctuations and it is intimately related to the entropy-driven 1/3 magnetization pseudo-plateau. As illustrated in Supporting Information Figure S6, the presence of pseudo-plateau plays an essential role in modulating the entropy landscape, thereby governing the magnetocaloric effect characteristics. Particularly, at relatively low temperatures (e.g., ~ 0.5 K), the magnetization $M$ increases upon heating above the 1/3 plateau but decreases below it, resulting in entropy changes $\delta S$ (related to $\delta M$) with opposite through the Maxwell relation. As shown in the inset of Supporting Information Figure S6, the integral around the plateau therefore cancels out, leading to a negligible $-\Delta S_M$ for field changes $\mu_0\Delta H \leq 0.8$ T. This behavior contrasts sharply with that at higher temperatures. For instance, at $T \geq 1.0$ K, $-\Delta S_M$ increases monotonically with $\mu_0\Delta H$ because the 1/3 magnetization pseudo-plateau disappears at further elevated temperatures. For systems without such magnetization pseudo-plateau nor Goldstone modes, the isentropic curves and magnetic Grüneisen ratio lines will rise up after decreasing the field below the dip position, e.g., as observed in the triangular-lattice dipolar magnet KBaGd(BO$_3$)$_2$.[25] Thus, unlike conventional ADR, where minimum temperatures occur at zero field, this persistent cooling effect extends over a finite field range. It makes ultra-low temperature measurements possible under finite field. This key difference makes the DSS materials particularly advantageous as low-temperature coolants. Our work thus demonstrates that the DSS systems, not only hosting exotic quantum spin states, but also function as exceptional sub-Kelvin coolants.

## 4. Experimental Methods

*Synthesis and powder diffraction*: Polycrystalline sample of EMSO was synthesized by a standard solid-state reaction method. Firstly, the appropriate amounts of the starting materials including Eu$_2$O$_3$ (99.99%), MgO (99.99%), and SiO$_2$ (99.99%) were homogeneously mixed and well ground in an agate mortar. Then, the mixture was pressed into a tablet with a diameter of 12.5 mm and placed in an alumina crucible. After that, the alumina crucible was calcined in a tube furnace filled with a reduced atmosphere (10 vol.%H$_2$} and 90 vol.%Ar). The temperature was slowly increased to 1050°C. After annealing for 12 h, the resultant product was cooled down to room temperature. Finally, the resultant product was reground and sintered to 1100°C for 30 h in the same condition to derive the ultimate product. The sample purity and crystal structure were determined at room temperature using a powder X-ray diffractometer (PXRD, model Bruker D8A A25) with Cu-K$\alpha$ radiation ($\lambda$ = 1.5418 Å). PXRD pattern was collected in the Bragg angle range of 10° ≤ 2$\theta$ ≤ 80° with a step size 0.02°. Refinement of the PXRD pattern was performed using GSAS software, and the schematic crystal structure was plotted by VESTA program.

*Magnetic and magnetocaloric measurements*: Magnetic measurements of polycrystalline EMSO above 2 K are carried out using a magnetic property measurement system (model MPMS-3). Magnetic measurements below 2 K are performed on a MPMS-3 with a $^3$He} insert. Isothermal magnetization measurements are performed at the selected temperatures under the magnetic field change of 0 to 7 T. The specific heat measurements are carried out using a Quantum Design Physical Property Measurement System (PPMS) equipped with a $^3$He-$^4$He Dilution Refrigerator insert. Measurements are carried out under applied magnetic fields of 0 T, 1 T, 2 T, and 4 T. The demagnetization cooling curves for various initial conditions are measured using a quasi-adiabatic two-layer magnetocaloric setup[25] integrated with a Quantum Design Physical Property Measurement System (PPMS). The outer layer consists of GGG single crystals, which acts as a thermal guard to minimize heat leaks from the PPMS chamber. For experimental measurements, we prepared a composite sample by thoroughly mixing 3.0 g of EMSO powder with an equal mass (3.0 g) of silver powder to enhance the overall thermal conductivity of the sample.

*Monte Carlo simulations*: For $S = 7/2$ spin system, we use the Markov-chain classical Monte Carlo (MC) approach to perform an importance sampling of the spin configuration, based on the Metropolis algorithm. The single spin update is used to study the DSS model for EMSO (Equation 1). In the MC simulations, we take the spins as unit vectors and set $J_{dp}$ as the energy unit. To fit the experimental magnetization curves, we perform simulations on lattice of 12×12 unit cells (each unit cell contains a four-site plaquette).

Without the long-range dipole-dipole interaction, the phase diagram of the original classical SS model has been investigated before:[3] for $J/J' > 1$, the ground state is a Néel AFM; while for $J/J' < 1$ it becomes a spiral state where the angle difference between NN spins is $\theta = \pi \pm \cos^{-1}(J/J')$. An complementary view of these magnetic states is provided by examining the static spin structure factor, which is defined as the Fourier transform of spin-spin correlation function $S(\mathbf{Q}) = \frac{1}{4 \times (N/4)^2} \sum_\eta \sum_{i,j \in \eta} \langle \mathbf{S}_i \cdot \mathbf{S}_j \rangle \exp(i\mathbf{Q}_i \cdot \mathbf{r}_{ij})$, where $\mathbf{r}_{ij} = \mathbf{r}_j - \mathbf{r}_i$ is the vector from site $i$ to site $j$, $N$ is the total number of sites, $\langle \cdot \rangle$ represents the average over different MC configurations, and $\eta$ is the sublattice index. The long-range magnetic order is quantified in terms of prominent peaks of $S(\mathbf{Q})$ in the momentum space. For the Néel AFM phase, the peak locates at $\Gamma = (0,0)$ [Note that the peak does not locate at $M = (\pi,\pi)$ since we calculate the correlation function between the same sublattice, and average over all four sublattices.] While for the spiral phase, there are peaks at $(0, \pm(\pi - \arccos(J/J')))$ or $(\pm(\pi - \arccos(J/J')), 0)$.

Incorporating long-range dipolar interactions using Ewald summation method,[28,39-41] with model parameters determined from magnetization data fitting, we find the static spin structure factor peak remains at the Γ point. In real space, the spin configuration is in an orthogonal phase shown in Figure 3b. Note this state is also found for dipolar square-lattice model with a relatively large dipolar interaction.[42,44] But different from the square lattice case, under magnetic fields, we find a 1/3 magnetization pseudo-plateau near 0.5 T at low temperature (and remains as a finite regime in the zero-temperature limit, as schematically shown in Figure 3b), and the corresponding real-space spin configuration is also presented in Figure 3b, which is an UUD state. The ratio of dipolar to Heisenberg coupling could stabilize diverse exotic spin states, which warrant a systematic theoretical investigation in future.[54]

**Author Contributions**

J.G., Z.M., W.L., and J.S. initiated this work. J.G., L.Z., X.H., L.T., Z.Y., H.X, X.G., and Z.L. performed the synthesis and magnetic measurements above 2 K. X.K. performed the magnetic measurements below 2 K. J.X. and X.L. performed the demagnetization cooling and specific heat measurements. J.W. and W.L. constructed the spin model and conducted theoretical analysis. J.W. conducted the MC simulations. All authors contributed to the analysis of the results and the preparation of the draft. Z.M., W.L., J.S., P.S., S.W., and B.S. supervised the project.

**Acknowledgements**

This work was financially supported by the National Science Foundation for Excellent Young Scholars (Grant Nos. 52222107 and 12222412), Strategic Priority Research Program of Chinese Academy of Sciences (Grant No. XDB1270000), National Key Projects for Research and Development of China (Grant No. 2024YFA1409200), and the Self-deployed Project of Ganjiang Innovation Research Institute, Chinese Academy of Sciences (Grant No. E355F001). We acknowledge the support from the Synergetic Extreme Condition User Facility (SECUF, https://cstr.cn/31123.02.SECUF).

**Competing Interest Statement**

The authors declare no competing interests.

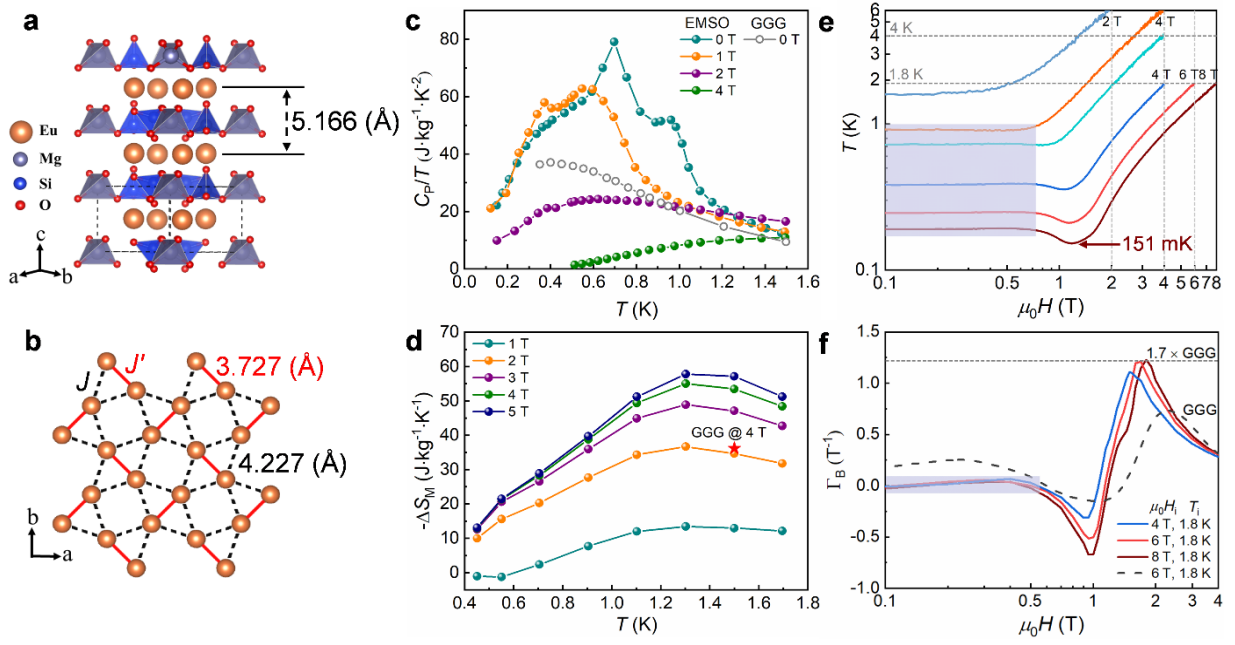

**Figure 1.** SS lattice structure and magnetothermal measurements on EMSO. The crystal structures of EMSO in the 3D lattice a) and the SS lattice of $Eu^{2+}$ ions within ab-plane b); c) The filled symbols correspond to the ultra-low temperature specific heat of EMSO in applied magnetic fields ranging from 0 to 4 T, and the open symbols correspond to the low-temperature zero-field specific heat of GGG; d) The entropy change $-\Delta S_M(T)$ curves for EMSO compound down to 400 mK. Entropy change of GGG is shown by a red star; e) The isentropic curves of EMSO at various initial conditions, measured in the quasi-adiabatic demagnetization process. A persistent cooling effect is observed, where the isentropes remain very flat below about 1 T, as indicated by the gray band; f) The derived magnetic Grüneisen ratio based on the measured isentropic lines. The dashed lines represent the commercial coolant GGG. The persistent cooling effect is also highlighted by the gray band.

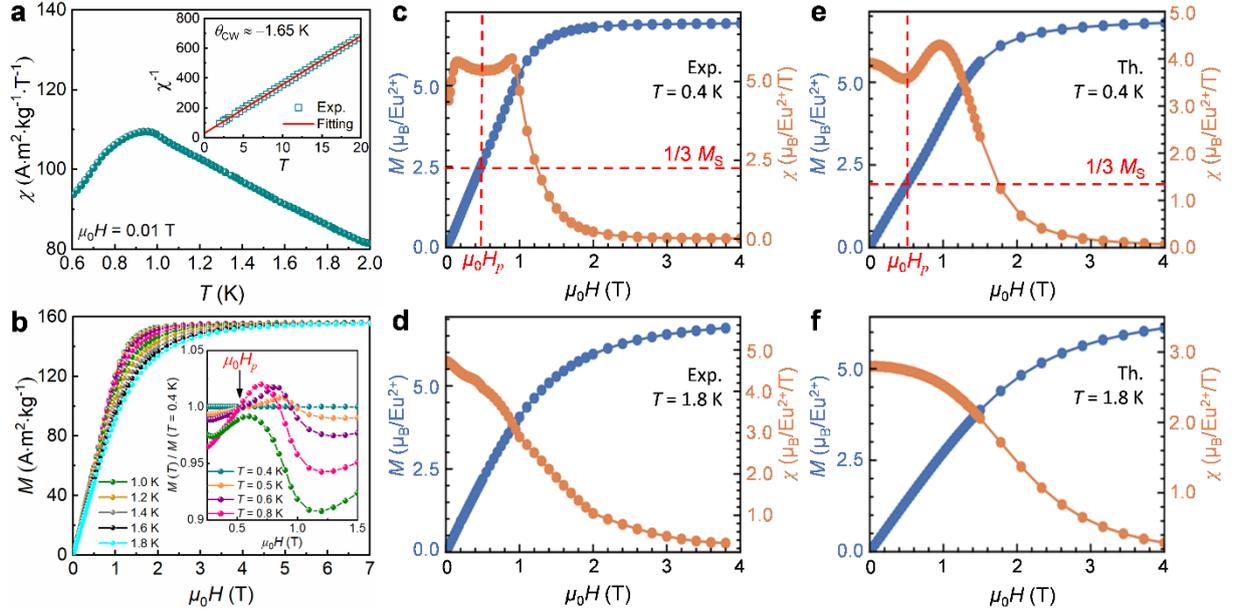

**Figure 2.** Magnetic susceptibility and magnetization curves. a) The measured magnetic susceptibility $\chi$ and its Curie-Weiss fitting (inset), from which we find the Curie-Weiss temperature $\theta_{CW} \approx -1.65$ K; b) Magnetization curves $M(\mu_0 H)$ measured at various temperatures are shown. The inset displays the normalized magnetization ratio $R_M \equiv M(T)/M(T = 0.4$ K), where the crossing point below $T \leq 0.8$ K signals the emergence of a 1/3 pseudo-plateau; c) and d) display the measured magnetization curves at 0.4 K (lowest temperature) and 1.8 K. At 0.4 K, the differential susceptibility $\chi = dM/d(\mu_0 H)$ exhibits a peak-dip-peak behavior with minimum near $\mu_0 H_p \approx 0.5$ T. This signature identifies the formation of a 1/3 magnetization pseudo-plateau state; e) and f) present simulated results from a DSS model, quantitatively reproducing the experimental observations — particularly the presence (0.4 K) and absence (1.8 K) of pseudo-plateau features.

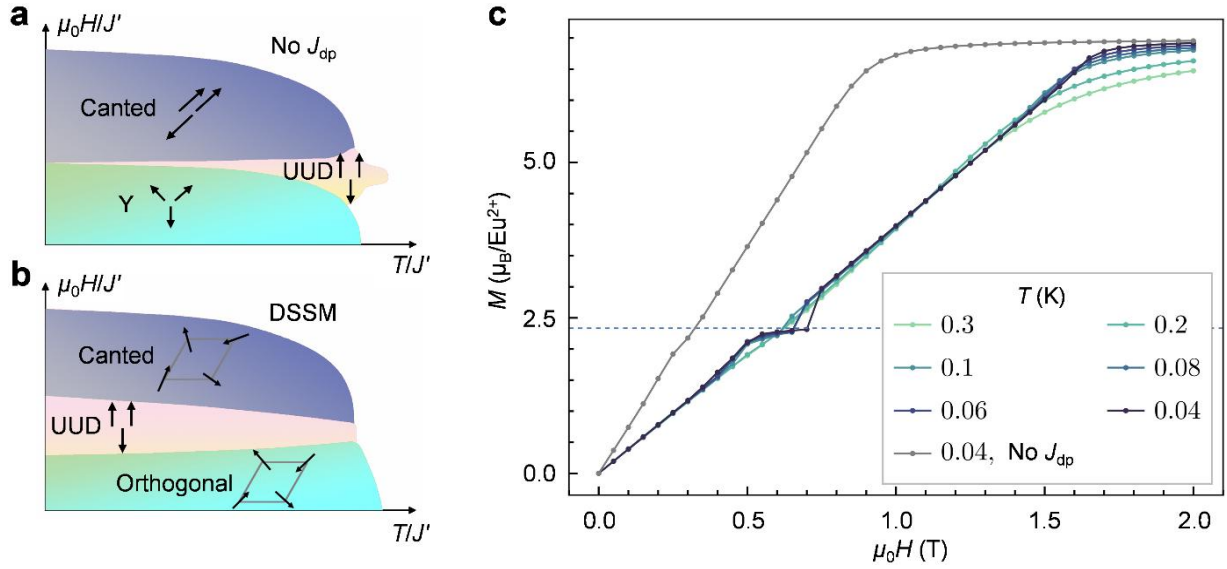

**Figure 3.** Phase diagram and 1/3 pseudo-plateau of the SS and DSS models. a) and b) Schematic phase diagram of the original SS model, and the case with dipolar interaction ($J_{dp}/J' = 0.59$), both near $J/J' = 1/2$; c) The magnetization process was simulated for models with and without dipolar interaction $J_{dp}$. The dipolar interaction results in a more pronounced magnetization plateau at low temperatures, which extends all the way to the ground state, with simulated results shown in b).

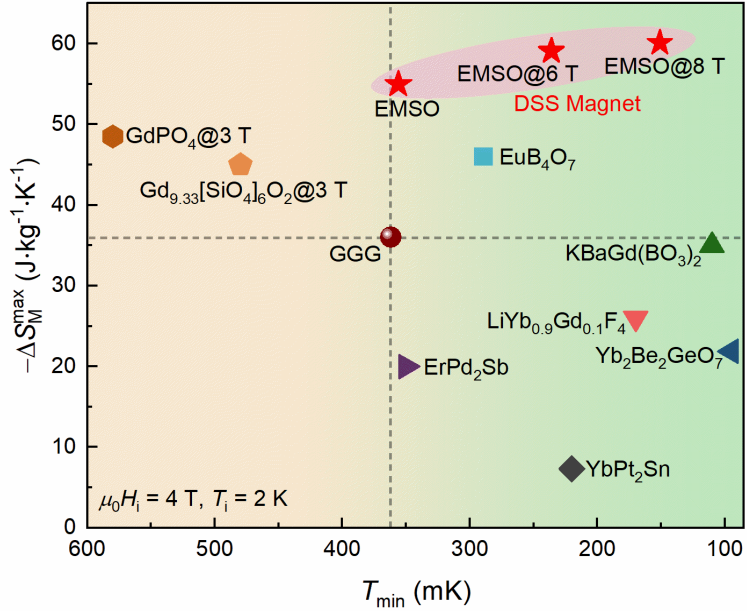

**Figure 4.** Comparison of magnetocaloric performance of sub-Kelvin ADR materials. The $-\Delta S_M^{max}$ is the maximum magnetic entropy change at $\mu_0 H$ = 0–4 T. $T_{min}$ is the attainable minimum temperature under initial conditions $\mu_0 H_i$ = 4 T and $T_i$ =2 K. Compared to commercial GGG, materials with large $-\Delta S_M^{max}$ value typically exhibit higher $T_{min}$, while those capable of achieving lower cooling temperatures generally possess smaller $-\Delta S_M^{max}$ value. This trade-off makes simultaneously achieving high cooling capacity and low cooling temperatures quite challenging. Notably, the high-spin DSS magnet EMSO overcomes this dilemma, thus demonstrating a superior magnetocaloric performance.

# Supporting Information for
Giant Magnetocaloric Effect in a High-Spin Shastry-Sutherland Dipolar Magnet

Jianjian Gong[1,2,*], Junsen Wang[3,4,*], Junsen Xiang[5,*], Zhaojun Mo[1,2,†], Lei Zhang[1], Xinyang Liu[4,5], Xuetong He[1,2], Lu Tian[1], Zhixing Ye[1], Huicai Xie[1], Xucai Kan[6], Xinqiang Gao[1,2], Zhenxing Li[7], Peijie Sun[5], Shouguo Wang[6], Wei Li[4,‡], Baogen Shen[1,2,5,6], and Jun Shen[1,2,7,§]

[1] Key Laboratory of Rare Earths, Ganjiang Innovation Academy, Chinese Academy of Sciences, Ganzhou 341119, China
[2] School of Rare Earths, University of Science and Technology of China, Hefei 230026, China
[3] Anhui Province Key Laboratory of Condensed Matter Physics at Extreme Conditions, High Magnetic Field Laboratory, Chinese Academy of Sciences, Hefei 230031, China
[4] Institute of Theoretical Physics, Chinese Academy of Sciences, Beijing 100190, China
[5] Beijing National Laboratory for Condensed Matter Physics, Institute of Physics, Chinese Academy of Sciences, Beijing 100190, China
[6] Anhui Provincial Key Laboratory of Magnetic Functional Materials and Devices, School of Materials Science and Engineering, Anhui University, Hefei 230601, China
[7] Department of Energy and Power Engineering, School of Mechanical Engineering, Beijing Institute of Technology, Beijing 100081, China

Jianjian Gong, Junsen Wang, and Junsen Xiang contributed equally to this work.

Corresponding Authors: Zhaojun Mo, Wei Li, Jun Shen.

**Section 1: Powder X-ray diffraction**

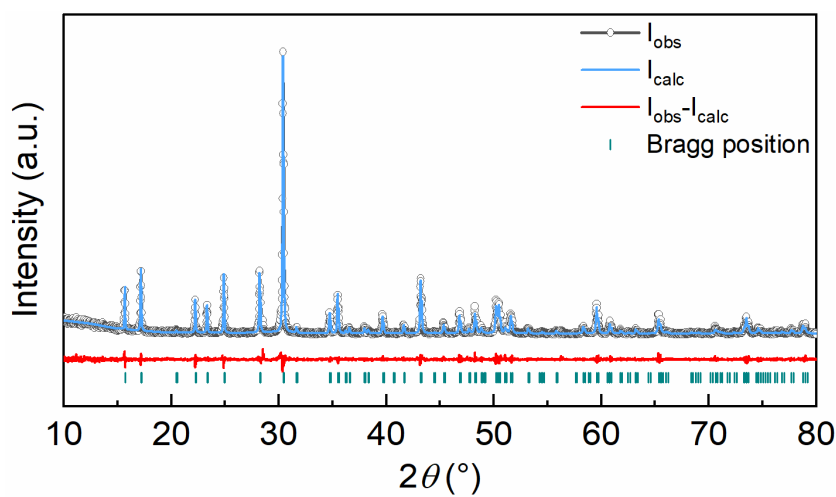

**Fig. S1.** Rietveld refinement of the room-temperature power X-ray diffraction (PXRD) data for the compound $Eu_2MgSi_2O_7$ (EMSO). Polycrystalline EMSO with light-green color was prepared using solid phase reaction method. The refined PXRD pattern shows the absence of impurity peaks and the phase purity of the sample. The highly coincident refinement results evidence that the sample crystallize in the tetragonal structure with the space group $P\bar{4}2_1m$. The refined parameters $R_p$, $R_{wp}$, and $\chi^2$ are reasonable and are 3.1%, 6.1%, and 1.3, respectively.

The obtained lattice parameters and volume are $a = b = 8.001$ Å, $c = 5.166$ Å, and $V = 330.747$ Å$^3$, respectively.

**Section 2: Magnetic and magnetocaloric measurements**

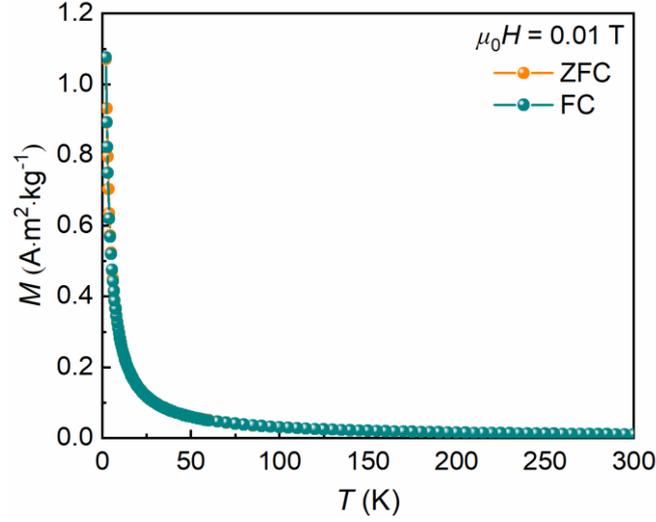

**Fig. S2.** Magnetization versus temperature of EMSO. Magnetization measured in the zero-field cooling (ZFC) and field cooling (FC) modes from 2 K to 300 K at an applied field of 0.01 T. The ZFC and FC curves exhibit a monotonically increasing trend with decreasing temperature, which indicates the paramagnetic (PM) behaviour above 2 K.

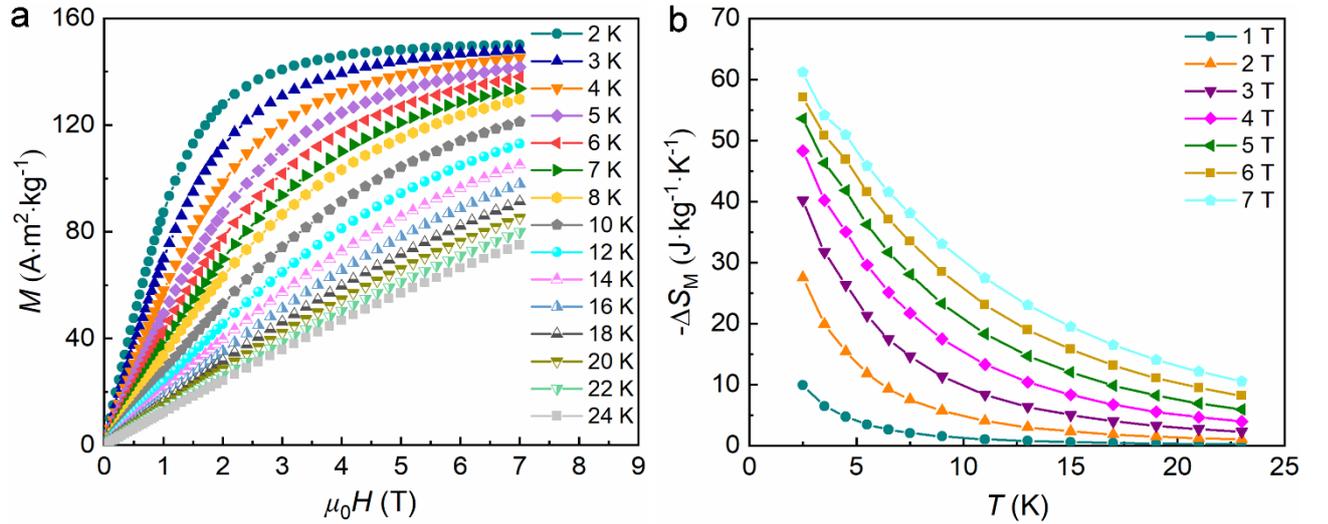

**Fig. S3.** Isothermal magnetization and magnetic entropy change above 2 K of EMSO. a) Isothermal magnetization curves measurements above 2 K; b) The magnetic entropy change ($-\Delta S_M$) is calculated using the Maxwell relation $\Delta S_M(T, \mu_0 \Delta H) = \int_0^{\mu_0 H} \left( \frac{\partial M}{\partial T} \right)_{\mu_0 H} \mu_0 dH$ from the $M(\mu_0 H)$ data. The $-\Delta S_M^{max}$ at 2.5 K is as high as 61.2 J·kg$^{-1}$·K$^{-1}$ under magnetic field change of 0–7 T.

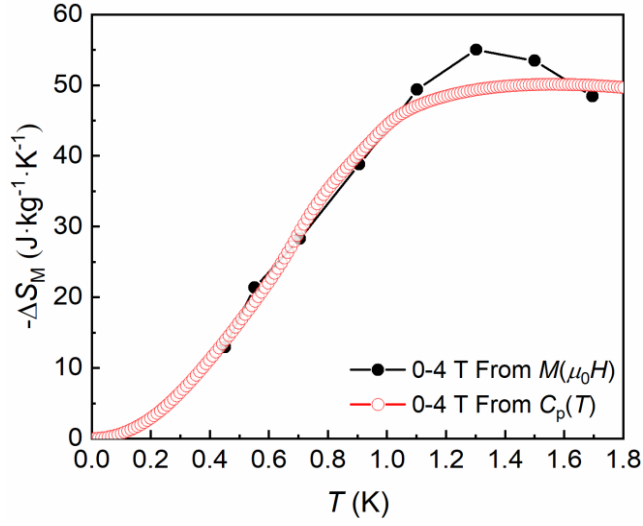

**Fig. S4.** Magnetic entropy change. The $-\Delta S_M(T)$ curves calculated from $C_p(T)$ (open symbols) and $M(\mu_0 H)$ (filled symbols) data. We extrapolated the zero-field $C_p$ values to $T = 0$ K based on the linear relationship between $C_p$ and $T^2$ below 0.3 K. Additionally, we assumed that the $C_p$ values under 4 T below 0.5 K are equal to zero. The $-\Delta S_M$ is calculated from $C_p(T)$ data using the relation $\Delta S_M(T, \mu_0 \Delta H) = \int_0^T \left( \frac{C(T, \mu_0 H) - C(T, 0)}{T} \right) dT$.

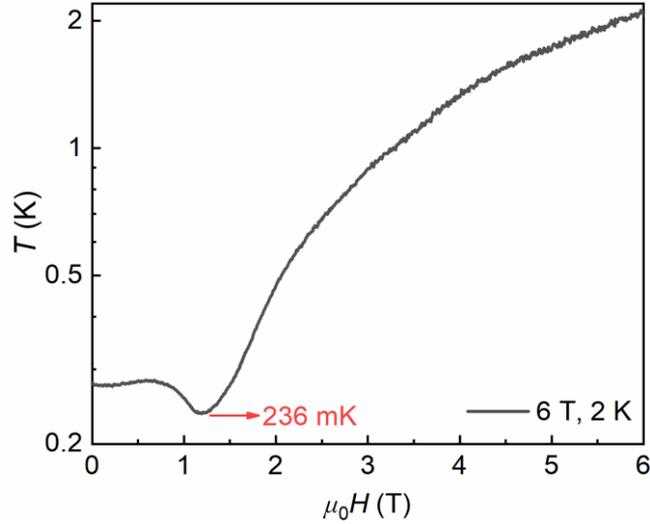

**Fig. S5.** Quasi-adiabatic demagnetization isentropes. The isentropic curve of EMSO at the initial conditions of $T_i = 2$ K and $\mu_0 H_i = 6$ T.

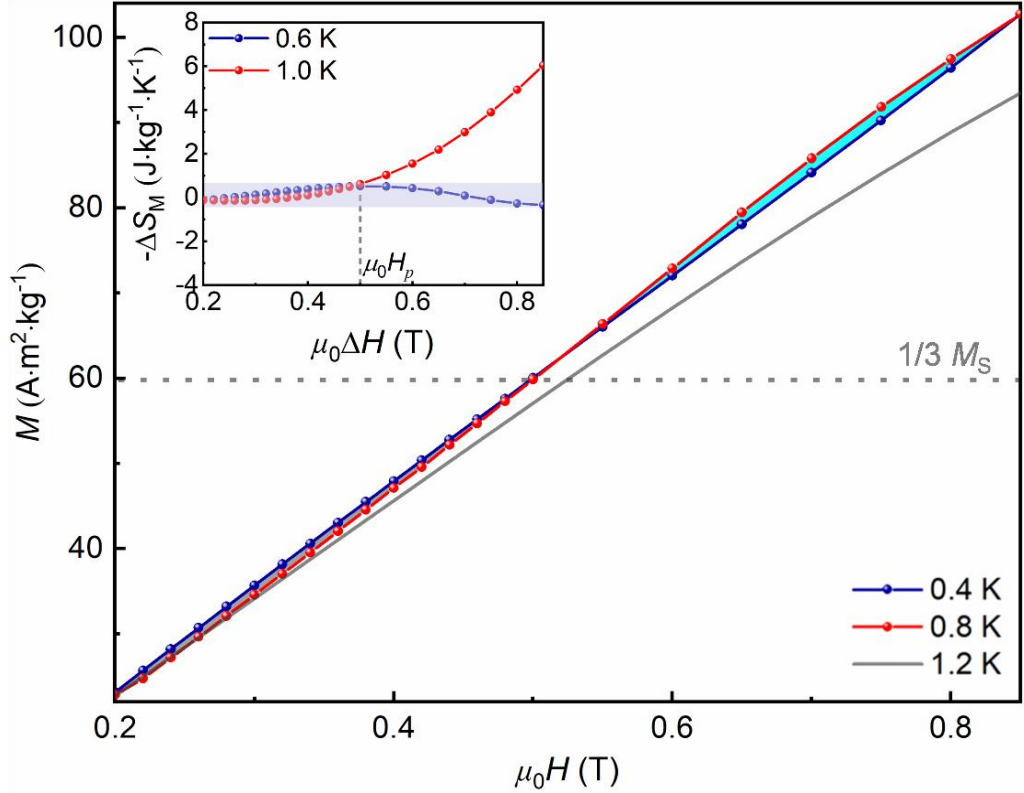

**Fig. S6.** Magnetization pseudo-plateau and persistent cooling. The magnetization curves at three different temperatures are shown, where a 1/3 pseudo-plateau can be recognized as the crossing point of the two low-temperature magnetization curves (indicated also by the horizontal dashed line). Two area enclosed by these two lines are colored in cyan and gray, corresponding to the entropy change with opposite signs. The inset shows the entropy change $-\Delta S_M$, which remains very small when the pseudo-plateau is present due to cancellation between positive and negative contributions; while increases monotonically as temperature is higher and the pseudo-plateau disappears. This results in small entropy change and stable low temperature below 1 T.